\documentclass[amsmath,amssymb,amsbsy,pra,twocolumn,preprintnumbers,showpacs]{revtex4-2}
\usepackage{graphicx,color}
\usepackage{dcolumn}
\usepackage{bm}
\usepackage{braket}
\usepackage{mathtools}
\usepackage{ulem}
\usepackage[breaklinks,colorlinks=true,linkcolor=red,urlcolor=blue,citecolor=magenta]{hyperref}

\begin{document}
\title{Square-root higher-order topological insulator on a decorated honeycomb lattice}
\author{Tomonari Mizoguchi}
\affiliation{Department of Physics, University of Tsukuba, Tsukuba, Ibaraki 305-8571, Japan}
\email{mizoguchi@rhodia.ph.tsukuba.ac.jp}
\author{Yoshihito Kuno}
\affiliation{Department of Physics, University of Tsukuba, Tsukuba, Ibaraki 305-8571, Japan}
\author{Yasuhiro Hatsugai}
\affiliation{Department of Physics, University of Tsukuba, Tsukuba, Ibaraki 305-8571, Japan}

\begin{abstract}
Square-root topological insulators are recently-proposed intriguing topological insulators,
where the topologically nontrivial nature of Bloch wave functions is inherited from the square of the Hamiltonian.
In this paper, we propose that higher-order topological insulators can also have their square-root descendants,
which we term square-root higher-order topological insulators. 
There, emergence of in-gap corner states is inherited from the squared Hamiltonian which hosts higher-order topology.
As an example of such systems, we investigate the tight-binding model on a decorated honeycomb lattice, 
whose squared Hamiltonian includes a breathing kagome-lattice model, a well-known example of higher-order topological insulators. 
We show that the in-gap corner states appear at finite energies, which coincides with the non-trivial bulk polarization. 
We further show that the existence of in-gap corner states results in characteristic single-particle dynamics, 
namely, setting the initial state to be localized at the corner, the particle stays at the corner even after a long time.
Such characteristic dynamics may experimentally be detectable in photonic crystals.
\end{abstract}
\maketitle

\section{Introduction}
Since the discovery of the quantum Hall effect~\cite{Klitzing1980} and finding of its topological origin~\cite{Thouless1982,Avron1983,Kohmoto1985},
topological phase of matter has attracted considerable interests in the field of condensed-matter physics. 
Topological insulators and superconductors (TIs and TSCs)~\cite{Hasan2010,Qi2011} 
are representatives of such systems of non-interacting fermions, 
where topologically nontrivial nature of the Bloch wave functions or Bogoliubov quasiparticles, characterized by topological indices~\cite{Schnyder2008,Kitaev2009,Ryu2010}, 
gives rise to robust gapless states at boundaries of samples~\cite{Kane2005,Fu2007}.
Such a relation between bulk topological indices and boundary states is called 
bulk-boundary correspondence~\cite{Hatsugai1993}.

It has also been revealed that incorporating crystalline symmetries~\cite{Hsiesh2012,Slager2012,Shiozaki2014,Ando2015,Po2017,Watanabe2018} and/or non-Hermiticity~\cite{Gong2018,Kawabata2019,Kawabata2019_2,Budich2019,Okugawa2019,Yoshida2019, Yoshida2019_2,Lee2019,Okuma2020,Yoshida2019_3} 
makes the topological phases more abundant.
Remarkably, the notion of bulk-boundary correspondence is enriched accordingly~\cite{Hashimoto2017,Hayashi2018,Benalcazar2017,Song2017,Schindler2018,Benalcazar2019,Khalaf2019,Yao2018,Kunst2018,Edvardsson2019,Yokomizo2019,Okuma2020,Yoshida2019_3}. 
Higher-order topological insulators (HOTIs)~\cite{Hashimoto2017,Hayashi2018,Benalcazar2017,Song2017,Schindler2018,Benalcazar2019,Ezawa2018,Xu2017,Kunst2018_2} 
are one of the examples exhibiting a novel type of bulk-boundary correspondence,
where topologically-protected boundary modes appear not at $d-1$ dimensional boundaries
but at $d-n$ ($n \geq 2$) dimensional boundaries, with $d$ being the spatial dimension of the bulk.
Nowadays, various theoretical models~\cite{Benalcazar2017,Schindler2018,Ezawa2018,Okugawa2019_2} as well as possible realizations in solid-states systems~\cite{Schindler2018_2,Ezawa2018_2,Kempkes2019,Sheng2019,Lee2020,Mizoguchi2019,Radha2020} 
have been proposed.
Furthermore, realization and potential applications in artificial materials~\cite{SerraGarcia2018,Imhof2018,Xue2019,Ni2019,Xie2018,Hassan2019,Ota2019,Zhang2019,Zhang2019_2,Wakao2020} 
have also been intensively pursued.
The effects of interactions~\cite{You2018,Dubinkin2019,Kudo2019,Araki2020} and disorders~\cite{Araki2019} have also been investigated.
\begin{figure}[b]
\begin{center}
\includegraphics[clip,width = 0.90\linewidth]{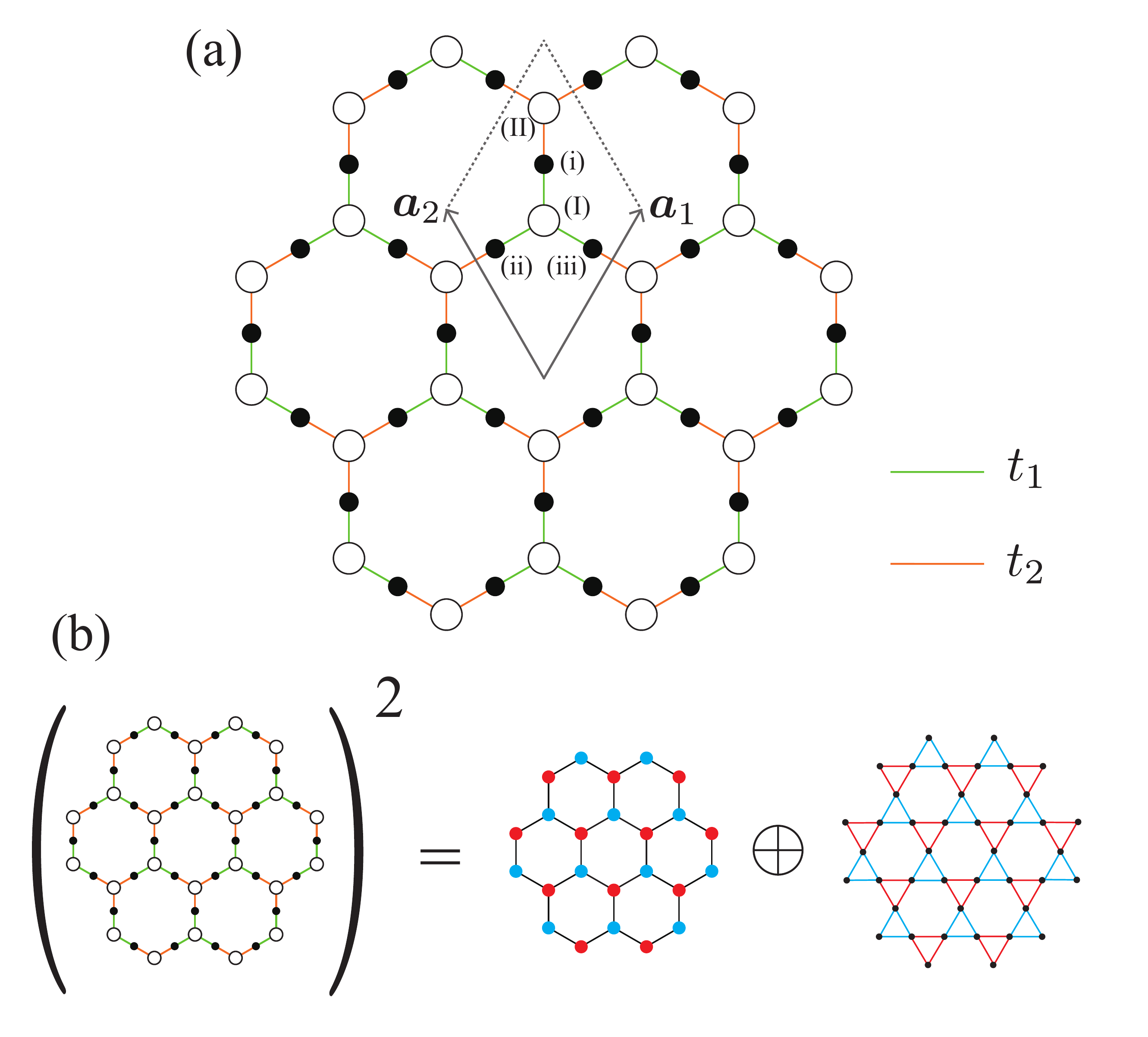}
\vspace{-10pt}
\caption{
(a) Model considered in this paper. 
Gray arrows represent the lattice vectors: 
$\bm{a}_1 = \left( \frac{1}{2}, \frac{\sqrt{3}}{2}\right)$ and $\bm{a}_2 = \left( -\frac{1}{2}, \frac{\sqrt{3}}{2}\right)$.
(b) A schematic figure of the relation between the original Hamiltonian and its square. 
The squared Hamiltonian is the direct sum of the honeycomb-lattice model with sublattice-dependent on-site potentials
and the breathing kagome-lattice model.}
  \label{fig:model}
 \end{center}
 \vspace{-10pt}
\end{figure}

Besides these developments, Arkinstall \textit{et al.} 
recently proposed a pathway to realize a novel type of TIs, by utilizing the square-root operation~\cite{Arkinstall2017}.
Namely, for a proper choice of positive-semidefinite tight-binding Hamiltonians, referred to as the parent Hamiltonians, 
their square-root Hamiltonian can be generated by inserting additional sublattice degrees of freedom.
Then the Hamiltonian thus obtained has a spectral symmetry around $E=0$, 
and the topologically protected boundary modes on the parent Hamiltonian are inherited to its square root.  
Such TIs are referred to as square-root TIs~\cite{remark_1}, and they are 
realized in, e.g., a diamond-chain photonic crystal~\cite{Kremer2020}.

Inspired by these findings, in the present paper, 
we propose that the idea of taking the square root can be applied to the HOTI.
Namely, we can construct the model of HOTI by taking the square-root of 
the well-known model of the HOTI; we term such systems square-root HOTIs.
The model thus obtained has characteristic corner states appearing at positive and negative energies
in a pairwise manner, which is a sharp contrast to the parent model.
To be concrete, we study a decorated honeycomb-lattice model.
Here a decorated honeycomb lattice stands for a honeycomb lattice with one additional site at each edge;
such a lattice structure is relevant to several solid-state systems, such as graphene superstructures~\cite{Shima1993},
metal-organic frameworks~\cite{Barreteau2017}, and 1T-TaS$_2$~\cite{Lee2019_2,Park2019},
and thus has long been studied mainly as an example of flat-band models.
As for the topological aspect, the possibility of the HOTI in a similar model was pointed out in Ref.~\onlinecite{Lee2019_2}, 
and here we present a viewpoint from its parent Hamiltonian. 
Namely, the parent Hamiltonian of the present system is 
the direct sum of the honeycomb-lattice model with different on-site potentials 
and the breathing kagome-lattice model~\cite{remark_bk}.
The breathing kagome-lattice model hosts the HOTI in which boundary modes are localized at the corner of the sample,
and this higher-order topology is succeeded to the decorated honeycomb model as well.
We demonstrate this by numerically calculating the eignevalues and eigenvectors to show the existence of the corner modes,
and by relating them to the topological invariant for both original and squared Hamiltonians. 

Besides the solid-state systems listed above, our model of the decorated honeycomb lattice 
is also experimentally feasible in photonic waveguide crystals, as recently reported in Refs.~\onlinecite{Yan2019,Tang2020}.
Therefore, to detect the topologically protected corner modes in the present system,
we calculate the single-particle dynamics. 
Such a single particle initial state can be easily prepared by injecting an excitation beam in photonic crystals, 
and its propagation pattern, 
corresponding to the particle dynamics of the quantum mechanical systems, is measurable. 
We find that, when the corner states exist, the particle localized at the corner sites stays 
at the same corner and does not spread into the bulk. 

The remainder of this paper is organized as follows. 
In Sec.~\ref{sec:model}, we explain the decorated honeycomb lattice model we use in this paper, and point out that 
its squared Hamiltonian corresponds to the direct sum of the honeycomb lattice model and the breathing kagome-lattice model.
We also present the exact dispersion relations. 
In Sec.~\ref{sec:corner}, we elucidate the higher-order topology of this model, 
by demonstrating the existence of the in-gap corner states in the finite system under the open boundary conditions.
We also calculate the topological invariant, i.e., the polarization, and show how the HOTI in the squared Hamiltonian is inherited to the original model. 
In Sec.~\ref{sec:dynamics}, we study the single-particle dynamics, and show its localized nature originating from the existence of the corner states. 
Finally, in Sec.~\ref{sec:summary}, we present a summary of this paper. 

\section{Model \label{sec:model}}
We consider the following tight-binding Hamiltonian on a decorated honeycomb lattice that has five sites per unit cell
[Fig.~\ref{fig:model}(a)]:
\begin{align}
H = \sum_{\bm{k}} \bm{C}^{\dagger}_{\bm{k}} \mathcal{H}_{\bm{k}} \bm{C}_{\bm{k}}, \label{eq:Ham_dh}
\end{align}
where $\bm{C}_{\bm{k}} = \left( C_{\bm{k},\bullet, (\mathrm{I})},C_{\bm{k},\bullet, (\mathrm{II}) },C_{\bm{k},\circ, (\mathrm{i}) },C_{\bm{k},\circ, (\mathrm{ii}) },C_{\bm{k},\circ, (\mathrm{iii})}
 \right)^{\rm T}$
and 
\begin{align}
\mathcal{H}_{\bm{k}} = 
\left(
\begin{array}{cc}
\mathcal{O}_{2,2} & \Phi_{\bm{k}}^\dagger  \\
\Phi_{\bm{k} } & \mathcal{O}_{3,3}\\
\end{array}
\right).
\end{align}
Here $\mathcal{O}_{l,m}$ represents the $l \times m$ zero matrix;
the $\Phi_{\bm{k}}$ is the $3\times 2$ matrix:
\begin{align}
\Phi_{\bm{k}} = \left(
\begin{array}{cc}
t_1 & t_2 \\
t_1 & t_2 e^{-i\bm{k} \cdot \bm{a}_1} \\
t_1 & t_2 e^{-i\bm{k} \cdot \bm{a}_2}\\
\end{array}
\right).
\end{align}
For the definitions of the lattice vectors $\bm{a}_1$ and $\bm{a}_2$, see Fig.~\ref{fig:model}(a).
Note that the model includes only nearest-neighbor (NN) hoppings with two different parameters $t_1$ and $t_2$.

The Hamiltonian is chiral-symmetric, i.e., 
$\mathcal{H}_{\bm{k}}$ satisfies $\gamma \mathcal{H}_{\bm{k}} \gamma = -\mathcal{H}_{\bm{k}}$
where
\begin{align}
\gamma = \left(
\begin{array}{cc}
I_{2,2} &  \mathcal{O}_{2,3}\\
 \mathcal{O}_{3,2}& -I_{3,3} \\
\end{array}
\right).
\end{align}
Here $I_{n,n}$ stands for the $n\times n$ identity matrix. 
This indicates the existence of the parent Hamiltonian whose square root corresponds to $\mathcal{H}_{\bm{k}}$~\cite{Arkinstall2017}.
The Hamiltonian also has $C_3$ symmetry around the sublattice (I),
\begin{align}
\mathcal{H}_{\left(C_3 \bm{k}\right)} = U_{\bm{k}}\mathcal{H}_{\bm{k}}U^{\dagger}_{\bm{k}},
\end{align}
where 
\begin{align}
U_{\bm{k}} = 
\left(
\begin{array}{ccccc}
1 & 0 & 0 & 0 & 0 \\
0 & e^{- i \bm{k}\cdot \bm{a}_2} & 0 & 0 & 0 \\
0 & 0 & 0 &0  & 1 \\
0&0&1&0 & 0\\
0&0&0&1&0\\
\end{array}
\right). \label{eq:c3}
\end{align}

The key feature of this Hamiltonian can be clarified by taking a square of the Hamiltonian matrix:
\begin{align}
\left[\mathcal{H}_{\bm{k}} \right]^2 = 
\left(
\begin{array}{cc}
h^{\rm (H)}_{\bm{k}} & \mathcal{O}_{2,3}  \\
\mathcal{O}_{3,2} &h^{\rm (K)}_{\bm{k}}  \\
\end{array}
\right),
\end{align}
where
\begin{subequations}
\begin{align}
h^{\rm (H)}_{\bm{k}} =\Phi^{\dagger}_{\bm{k}}  \Phi_{\bm{k}} ,
\end{align}
is the $2 \times 2$ matrix, and 
\begin{align}
h^{\rm (K)}_{\bm{k}} = \Phi_{\bm{k}} \Phi^{\dagger}_{\bm{k}},  \label{eq:Ham_k}
\end{align}
\end{subequations}
is the $3 \times 3$ matrix.
Remarkably, $h^{\rm (H)}_{\bm{k}}$ 
equals to the Hamiltonian of the honeycomb-lattice model with different on-site potentials on two sublattices 
[$3t_1^2$ for (I) and $3t_2^2$ for (II)],
while $h^{\rm (K)}_{\bm{k}}$ 
equals to the Hamiltonian of the breathing kagome-lattice model, as schematically depicted in Fig.~\ref{fig:model}(b).
From the real-space viewpoint, this is understood as follows.
The particle on a white (black) site can move to the neighboring black (white) sites by operating the Hamiltonian.
Then, operating the Hamiltonian twice corresponds to \textit{two NN moves} 
of the particles~\cite{Attig2017,Mizoguchi2019_2}, meaning that a particle on 
a white (black) site can either go to the neighboring white (black) sites or come back to the original site;
the former networks the honeycomb (kagome) lattice formed by white (black) sites, while the latter corresponds to the on-site potentials. 
We call the squared Hamiltonian, 
$\left[\mathcal{H}_{\bm{k}} \right]^2$, the parent Hamiltonian of this model.
The role of the parent Hamiltonian here is to help us understand the topological characters
of its child, namely the decorated honeycomb model of Eq.~(\ref{eq:Ham_dh}).
As such, the fine-tuned nature of the parent Hamiltonian (i.e., the on-site potential is not independent of the NN hoppings) does not matter to the following discussions, as the main target we deal with is not the parent Hamiltonian itself but its child.

As is well-known, the HOTI is realized in the breathing kagome-lattice model~\cite{Ezawa2018,Xu2017,Kunst2018},
and this higher-order topology of the parent Hamiltonian is indeed succeeded to its descendant $\mathcal{H}_{\bm{k}}$, as we will show later.
We note that the uniform on-site potential in the kagome-lattice Hamiltonian in Eq.~(\ref{eq:Ham_k}) is an artifact of the 
square operation, and it does not affect the topological nature.
\begin{figure}[t]
\begin{center}
\includegraphics[clip,width = 0.90\linewidth]{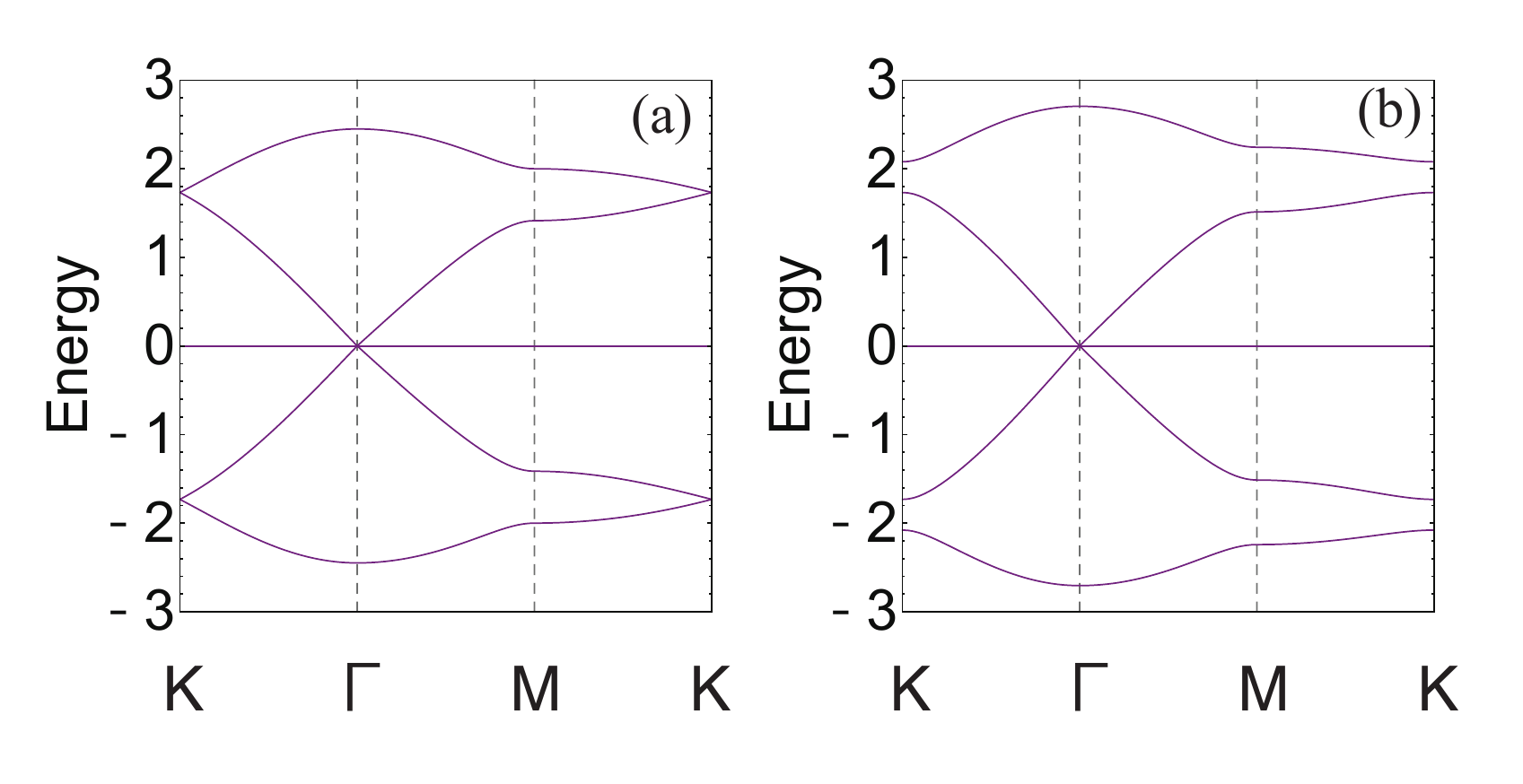}
\vspace{-10pt}
\caption{Band structures of the Hamiltonian $\mathcal{H}_{\bm{k}}$ for (a) $t_1 = 1$, $t_2 = 1$, and (b) $t_1=1$, $t_2 = 1.2$.
$\Gamma = (0,0)$, K$=\left(\frac{4\pi}{3},0\right)$, and M=$\left(\pi, \frac{\pi}{\sqrt{3}} \right)$ are high-symmetry points in the Brillouin zone. }
  \label{fig:band}
 \end{center}
 \vspace{-10pt}
\end{figure}

By utilizing the square of the Hamiltonian,
the dispersion relations of five bands can be obtained, because 
the dispersion relations can be obtained for both the honeycomb and the breathing-kagome models~\cite{Hatsugai2011}. 
To be concrete, the dispersion relations for $h^{\rm (H)}_{\bm{k}}$ are
\begin{eqnarray}
\varepsilon^{\rm (H)}_{\bm{k}} = E^{\pm}_{\bm{k}}  = \frac{3}{2} \left[t_1^2  +t_2^2 \pm \sqrt{\left( t_1^2-t_2^2 \right)^2 + 4t_1^2 t_2^2 |\Delta(\bm{k})|^2} \right], \nonumber \\  \label{eq:dispH}
\end{eqnarray}
and those for $h^{\rm (K)}_{\bm{k}}$ are
\begin{eqnarray}
\varepsilon^{\rm (K)}_{\bm{k}} = 0, E^{\pm}_{\bm{k}},  \nonumber \\ \label{eq:dispK}
\end{eqnarray}
where $\Delta(\bm{k}) = \left(1 + e^{i\bm{k}\cdot \bm{a}_1} + e^{i\bm{k}\cdot \bm{a}_2}\right)/3$.
We note that the dispersion relations for $\varepsilon^{\rm (H)}_{\bm{k}}$ and for $h^{\rm (K)}_{\bm{k}}$ are the same except for the existence of the zero-energy flat band 
for $h^{\rm (K)}_{\bm{k}}$. 
It follows from Eqs. (\ref{eq:dispH}) and (\ref{eq:dispK}) that 
the dispersion relations of the decorated honeycomb-lattice models are
\begin{align}
\varepsilon_{\bm{k}} = 0,\pm \sqrt{E^{\pm}_{\bm{k}} }.
\end{align}
The band structures for several parameters are depicted in Fig.~\ref{fig:band}.
There is a flat band at zero energy regardless of the choice of parameters, which originates from the chiral symmetry with sublattice imbalance from the conventional viewpoint~\cite{Lieb1989,Sutherland1986,Brouwer2002,Mizoguchi2019_3}. 
Alternatively, we can regard that this flat band is inherited from the breathing kagome bands of the squared Hamiltonian. 
When $|t_1| = |t_2|$, the squared Hamiltonian equals that for the honeycomb-lattice model with a \textit{uniform} on-site potential
plus the \textit{isotropic} kagome-lattice model, 
thus the Dirac cones appear at K and K$^\prime$ points [Fig.~\ref{fig:band}(a)],
whereas they are gapped out when $|t_1| \neq |t_2|$ [Fig.~\ref{fig:band}(b)].
This gap opening makes it possible to seek the HOTI in this model.

We further point out the eigenvectors of $\mathcal{H}_{\bm{k}}$ can be constructed from those of $h^{\rm (H)}_{\bm{k}}$ and $h^{\rm (K)}_{\bm{k}}$.
To be specific, let us focus on the first and the fifth bands, which are relevant to the higher-order topology discussed in the next section.
Let $\bm{u}^{\rm (H)}(\bm{k})$ be the eigenvector of $h^{\rm (H)}_{\bm{k}}$ with the eigenenergy $E^+_{\bm{k}}$
and $\bm{u}^{\rm (K)}(\bm{k})$ be that of $h^{\rm (K)}_{\bm{k}}$.
In fact, $\bm{u}^{\rm (K)}(\bm{k})$ can be written as~\cite{Mizoguchi2019_2,Hatsugai2011}
\begin{eqnarray}
\bm{u}^{\rm (K)}(\bm{k}) = \Phi_{\bm{k}} \bm{u}^{\rm (H)}(\bm{k}).
\end{eqnarray}
Note that $\bm{u}^{\rm (H)}(\bm{k})$ and $\bm{u}^{\rm (K)}(\bm{k})$ are not necessarily normalized. 
Then, it follows that the fifth eigenvector of $\mathcal{H}_{\bm{k}}$, which we write $\bm{u}_5(\bm{k})$,
can be written as 
\begin{eqnarray}
\bm{u}_5(\bm{k}) = \frac{1}{\mathcal{N}_{\bm{k}}}
\left(
\begin{array}{c}
\sqrt{E^{+}_{\bm{k}}} \bm{u}^{\rm (H)}(\bm{k}) \\
\bm{u}^{\rm (K)}(\bm{k}) \\
\end{array}
\right), \label{eq:u5}
\end{eqnarray}
with $\mathcal{N}_{\bm{k}}$ being the normalization constant. 
One can easily check that $\bm{u}_5(\bm{k})$ is 
indeed the eigenvector of $\mathcal{H}_{\bm{k}}$ with 
the eigenenergy $\sqrt{E^+_{\bm{k}}}$, as
\begin{eqnarray}
\mathcal{H}_{\bm{k}} \bm{u}_5(\bm{k})  &=&
\left(
\begin{array}{cc}
\mathcal{O}_{2,2} & \Phi_{\bm{k}}^\dagger  \\
\Phi_{\bm{k} } & \mathcal{O}_{3,3}\\
\end{array}
\right) \bm{u}_5(\bm{k}) \nonumber \\
&=& \frac{1}{\mathcal{N}_{\bm{k}}} 
\left(
\begin{array}{c}
\Phi^\dagger_{\bm{k}} \bm{u}^{\rm (K)} (\bm{k}) \\
\sqrt{E^{+}_{\bm{k}} } \Phi_{\bm{k}} \bm{u}^{\rm (H)} (\bm{k}) \\
\end{array}
\right)\nonumber \\
&=& 
\frac{1}{\mathcal{N}_{\bm{k}}} 
\left(
\begin{array}{c}
\Phi^\dagger_{\bm{k}} \Phi_{\bm{k}} \bm{u}^{\rm (H)} (\bm{k}) \\
\sqrt{E^{+}_{\bm{k}} } \bm{u}^{\rm (K)} (\bm{k}) \\
\end{array}
\right)\nonumber \\
&=& 
\frac{1}{\mathcal{N}_{\bm{k}}} 
\left(
\begin{array}{c}
E^{+}_{\bm{k}} \bm{u}^{\rm (H)} (\bm{k}) \\
\sqrt{E^{+}_{\bm{k}} } \bm{u}^{\rm (K)} (\bm{k}) \\
\end{array}
\right)\nonumber \\
&=& \sqrt{E^{+}_{\bm{k}} } \bm{u}_5 (\bm{k}).
\end{eqnarray}
Further, due to the chiral symmetry, the first eigenvector $\bm{u}_1 (\bm{k})$ can be obtained as
\begin{eqnarray}
\bm{u}_1(\bm{k}) 
= \gamma \bm{u}_5(\bm{k}) 
=
\frac{1}{\mathcal{N}_{\bm{k}}}
\left(
\begin{array}{c}
\sqrt{E^{+}_{\bm{k}}} \bm{u}^{\rm (H)}(\bm{k}) \\
- \bm{u}^{\rm (K)}(\bm{k}) \\
\end{array}
\right). \label{eq:u1}
\end{eqnarray}

\section{Corner states and their topological origin \label{sec:corner}}
\begin{figure}[b]
\begin{center}
\includegraphics[clip,width = 0.90\linewidth]{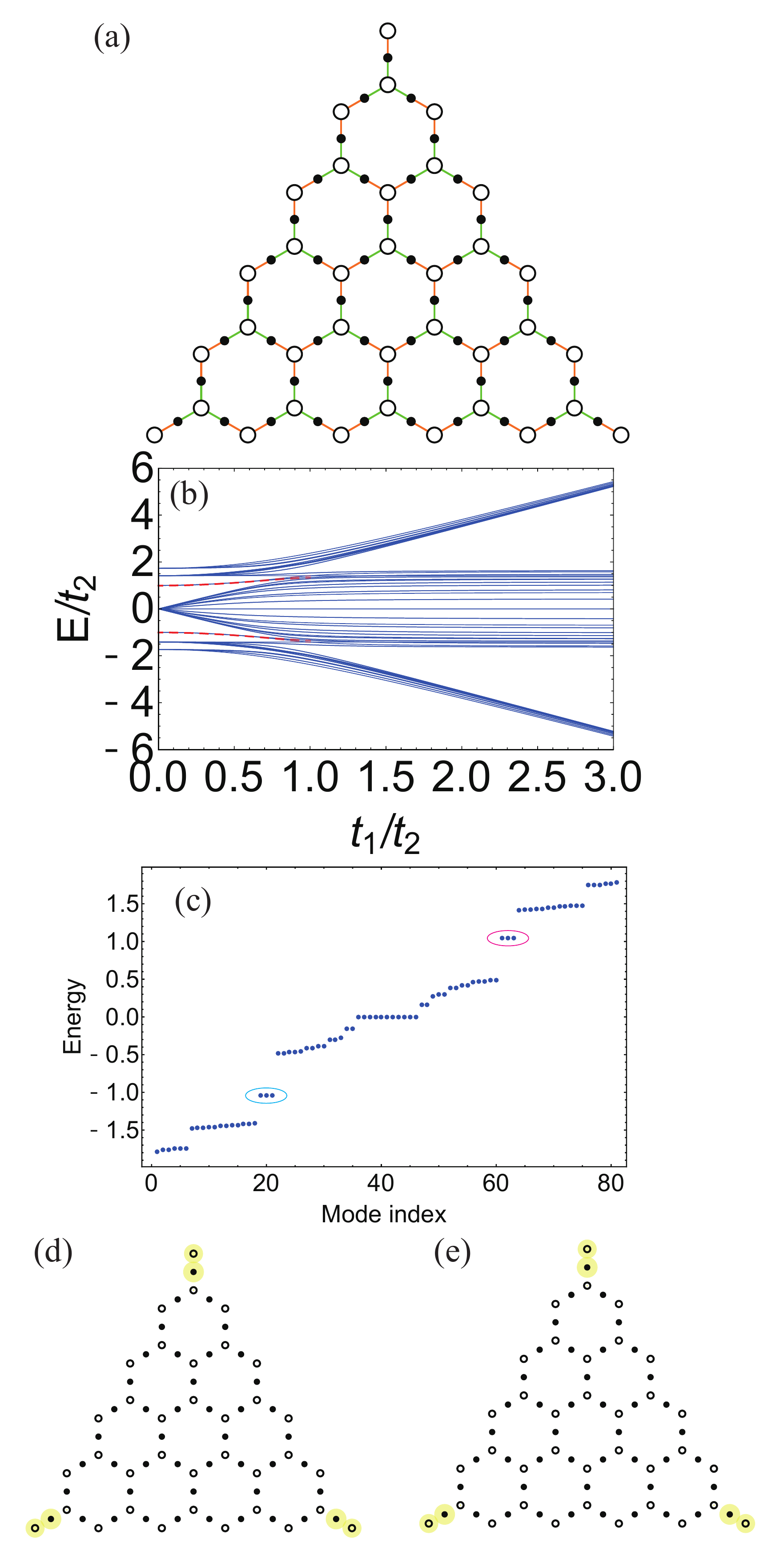}
\vspace{-10pt}
\caption{(a) Finite system under the open boundary conditions, consisting of 45 black and 36 white sites.  
(b) The energy spectrum as a function of $t_1/t_2$. 
The lines colored in red correspond to the in-gap corner states. 
(c) The energy spectrum for $t_1=0.3$, $t_2 = 1$. 
The in-gap corner states are encircled by cyan (lower) and magenta (upper) ellipses.
The charge density of (d) the lower three in-gap states
and (e) the higher three in-gap states. The radii of yellow shaded dots represent 
the probability density. }
  \label{fig:open}
 \end{center}
 \vspace{-10pt}
\end{figure}
In this section, we discuss the topological nature of the present model, focusing on 
the higher-order topological phase and its relation to the squared Hamiltonian. 

We first study the finite sample shaped into triangle under the open boundary conditions, shown in Fig.~\ref{fig:open}(a).
For this system, the energy spectrum as a function of $t_1/t_2$ is plotted in Fig.~\ref{fig:open}(b); 
without loss of generality, we set $t_{1/2} \geq 0$ hereafter~\cite{remark2}. 
We see there exist in-gap states for $t_1 /t_2 < 1$, whereas they vanish for $t_1 / t_2 > 1$;
as we have seen in the previous section, $t_1/ t_2 = 1$ corresponds to the gap-closing point, indicating the topological phase transition at this point. 
Figure~\ref{fig:open}(c) is the energy spectrum for fixed values of $t_1$ and $t_2$ with $t_1/t_2 < 1$.
We see that each in-gap state has three-fold degeneracy, as indicated by cyan and magenta ellipses. 

To further reveal the real-space profiles of the in-gap states, 
we plot the probability distribution for the in-gap states:
\begin{align}
N^{\nu}_{j}  = |\phi^{(\nu)}_{j}|^2,
\end{align}
where $j$ denotes the sites and $\phi^{(\nu)}_{j}$ is the real-space wave function defined 
for the eigenmode $\gamma_\nu$ as 
\begin{align}
\gamma_\nu = \sum_j \left(\phi^{(\nu)}_{j} \right)^{\ast} C_j,
\end{align}
with $C_j$ being the annihilation operator at the site $j$.
The results are shown in Fig.~\ref{fig:open}(d) for the negative-energy modes and Fig.~\ref{fig:open}(e) for the positive-energy modes~\cite{remark3}, respectively;
here we take averages over the three quasi-degenerate states.
We clearly see that the ing-gap states are indeed the corner states, manifesting that the HOTI is realized in the present model. 
It is worth noting that the corner states have large amplitude on both black sites and white sites at the corners. 
In fact, the proper choice of the corner-site termination is necessary to obtain the in-gap corner states,
which is a ubiquitous feature of two-dimensional HOTIs.
The present corner termination is chosen such that 
the square of the Hamiltonian of the finite system corresponds to 
the triangular geometry for the breathing kagome lattice which was used in Ref.~\onlinecite{Ezawa2018}.
\begin{figure}[t]
\begin{center}
\includegraphics[clip,width = 0.98\linewidth]{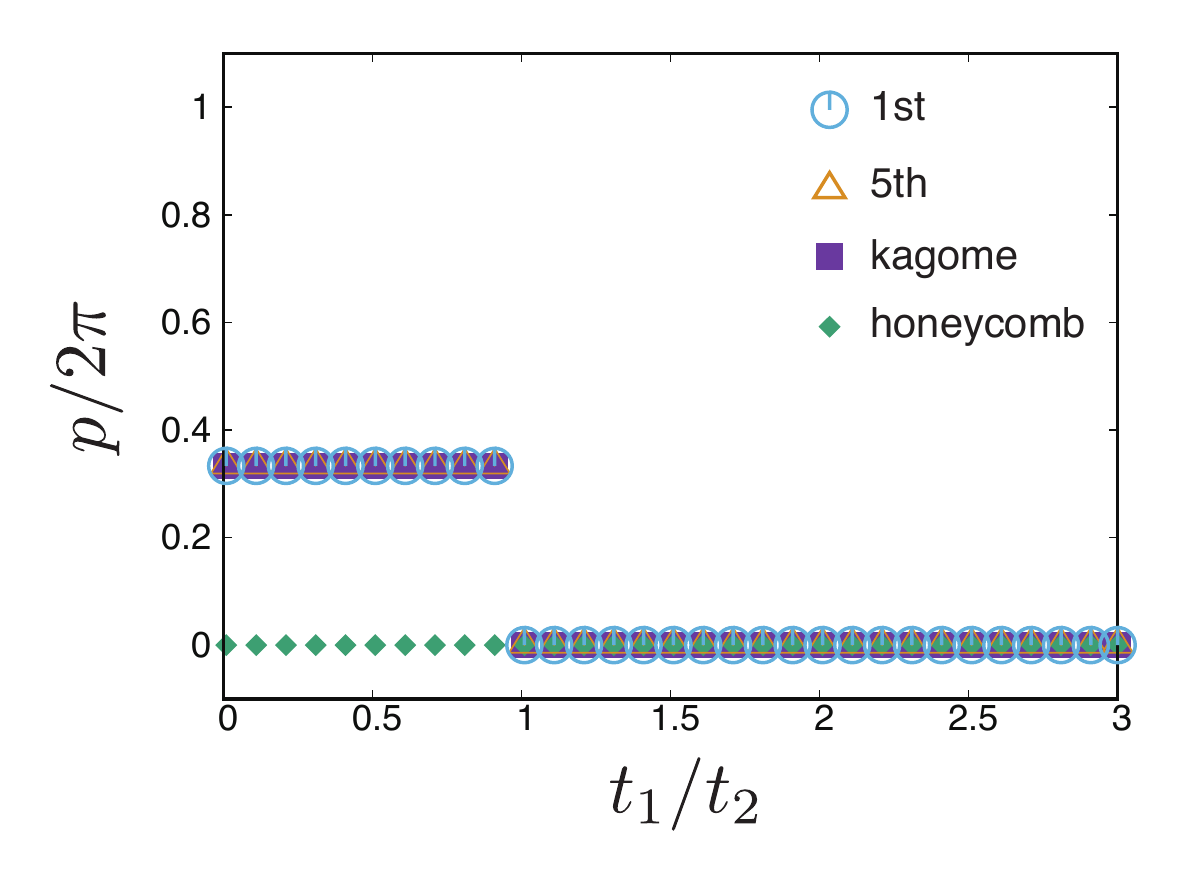}
\vspace{-10pt}
\caption{Polarization of Eq.~(\ref{eq:pol}) as a function of $t_1/t_2$. 
Blue dots and orange triangles are for the first and the fifth bands of the decorated honeycomb-lattice model, respectively.
Purple squares are for the upper dispersive band of the breaking kagome-lattice model $h^{\rm (K)}_{\bm{k}}$, 
and the green diamonds are for the upper dispersive band of the honeycomb-lattice model $h^{\rm (H)}_{\bm{k}}$. }
  \label{fig:pol}
 \end{center}
 \vspace{-10pt}
\end{figure}
\begin{figure*}[t]
\begin{center}
\includegraphics[clip,width = 0.95\linewidth]{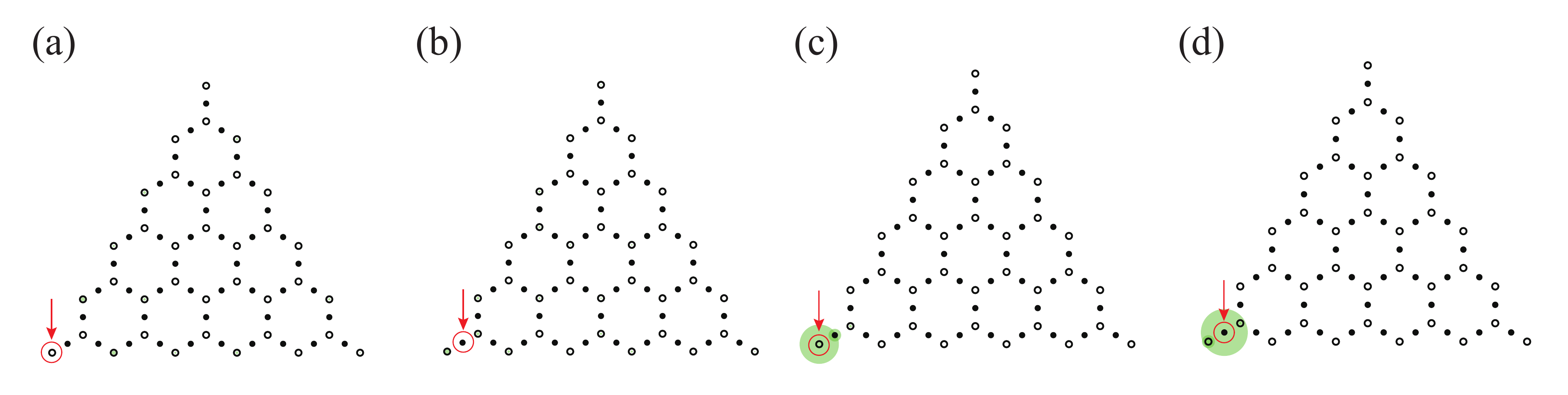}
\vspace{-10pt}
\caption{Charge density $N_{\ell}(t)$ of Eq.~(\ref{eq:chargedensity}) at $t=500$
represented by the radii of green shaded dots. 
The panels (a) and (b) are for $t_1=1$, $t_2=0.3$, 
and (c) and (d) are for 
$t_1=0.3$, $t_2=1$.
Red circles represent the initial position of the particle.
For (a) and (b), the green dots are smaller than black and white symbols representing the sites.}
  \label{fig:dynamics}
 \end{center}
 \vspace{-10pt}
\end{figure*}

Next, we discuss the topological origin of the corner states. 
To this end, we employ the polarization as a topological invariant.
For $C_3$-symmetric systems, it can be written in a concise form~\cite{Benalcazar2019,Ni2019,Fang2012}:
\begin{align}
2\pi p_n \equiv  \arg \theta_n(\mathrm{K})  \hspace{1mm} (\mathrm{mod} \hspace{0.5mm} 2\pi), \label{eq:pol}
\end{align}
where 
\begin{align}
\theta_n (\bm{k}) =\bm{u}^\dagger_n(\bm{k}) \cdot \left(U_{\bm{k}}\bm{u}_n (\bm{k}) \right).
\end{align}
The $C_3$ symmetry enforces the quantization of $p_n$ as 
$p_n = \frac{l}{3} \cdot l $ with $l = 0,1,2$. 
To characterize the corner states, we focus on the first and the fifth bands.
In Fig.~\ref{fig:pol}, we plot the polarization. 
Clearly, both $p_1$ and $p_5$ take $\frac{1}{3}$ for $t_1/t_2 < 1$, where the corner states emerge,
and $0$ for $t_1/t_2 >1$, where the corner states do not emerge.
The jump of $p_n$ occurs at $t_1 = t_2$, where the band gap closes and the topological phase transition occurs. 

How is the polarization related to the squared Hamiltonian? 
To see this, we calculate the polarization 
for $\bm{u}^{\rm (H)} (\bm{k}) $ and $\bm{u}^{\rm (K)} (\bm{k})$. 
Note that the $C_3$ rotation matrix $U_{\bm{k}}$ is block-diagonalized into the kagome sector and the honeycomb sector [see Eq.~(\ref{eq:c3})],
thus in the calculation of $p_{\rm K/H}$ we employ each sector of $U_{\bm{k}}$ as a $C_3$ rotation matrix. 
The results are plotted in Fig.~\ref{fig:pol}.
We see that $p$ changes from 
$\frac{1}{3}$ ($t_1/t_2<1$) to $0$ ($t_1/t_2>1$) for $h^{\rm (K)}_{\bm{k}}$, 
whereas it is $0$ for $h^{\rm (H)}_{\bm{k}}$.
This indicates that the higher-order topology of the present model is inherited from the breathing kagome-lattice model.
However, as we have pointed out, the corner modes have amplitudes on both honeycomb (white) and kagome (black) sites,
meaning that the actual corner states are not described by the kagome lattice alone. 
One can also find that the following relation of the topological invariants between the original model and the squared model holds:
\begin{align}
2\pi p_1 \equiv 2\pi  p_5 \equiv 2\pi ( p_{\rm K} + p_{\rm H} ) \hspace{1mm} (\mathrm{mod} \hspace{0.5mm} 2\pi), \label{eq:sumrule}
\end{align} 
which follows from Eqs.~(\ref{eq:u5}) and (\ref{eq:u1})~\cite{remark_4}.
Note that, 
unlike the case of the square-root TI in the diamond chain~\cite{Kremer2020},
the fractionalization of the topological invariant does not occur in the present model,
because the key symmetries are not broken by the square-root operation.

\section{Dynamical properties \label{sec:dynamics}}
In this section, we address the dynamics of the single-particle state associated with the localized corner states.
Although the following formulation is written in the language of the second quantization, 
one can employ the same scheme to describe the dynamics of electro-magnetic waves in photonic crystals~\cite{Arkinstall2017,Kremer2020,Rechtsman2013,Mukherjee2018,Ozawa2019,Longhi2019}.
To be concrete, time dependence of single-particle wave functions in tight-binding models 
can be translated into the $z$ dependence of electro-magnetic waves in photonic crystals, where $z$ stands for the spatial direction in which the wave  propagates.
Thus, the characteristic dynamics we show below may experimentally be realizable by using the photonic crystal setup.

Suppose that the single particle is localized at the site $j$ in the initial state: 
\begin{align}
\ket{\Psi(0)}  = C^\dagger_j \ket{0},
\end{align}
where $\ket{0}$ represents the vacuum state. 
Then, the wave function at time $t$ can be obtained by the unitary time evolution:
\begin{align}
\ket{\Psi(t)} = e^{-i Ht} \ket{\Psi(0)} = \sum_{\nu} \left(\phi^{(\nu)}_{j} \right)^{\ast} e^{-i \varepsilon_\nu t} \gamma^{\dagger}_\nu \ket{0},
\end{align} 
where $\varepsilon_\nu$ denotes the eigenenergy of $\nu$-th mode,
and we set $\hbar = 1$.
Thus the density at the site $\ell$ is obtained as
\begin{align}
N_{\ell} (t) = & \bra{\Psi(t)} C_\ell^\dagger C_\ell \ket{\Psi(t)} \notag \\
= & \left| \sum_{\nu} e^{- i \varepsilon_\nu t}
\left(\phi_j^{(\nu)}\right)^\ast \phi_\ell^{(\nu)}  \right|^2.  \label{eq:chargedensity}
\end{align}
We emphasize that this quantity can be easily observed in the photonic crystal. In Fig.~\ref{fig:dynamics}, 
we plot $N_{\ell}(t)$ for large $t$ (compared with the band width of the system).
Here the initial state is chosen such that the particle is localized at either the white site or the black site on the left-bottom corner.
In the case without the corner states [Figs.~\ref{fig:dynamics}(a) and ~\ref{fig:dynamics}(b)], the particle spreads over the entire system, 
thus the amplitude at individual sites becomes small.
In contrast, in the case with the corner states [Figs.~\ref{fig:dynamics}(c) and ~\ref{fig:dynamics}(d)],
the particle stays at the left-bottom corner even after a long time.

\section{Summary \label{sec:summary} }
To summarize, we have proposed the HOTI analog of the square-root TIs,
which we term the square-root HOTI. 
We study the decorated honeycomb-lattice model as an example, which can be regarded as a 
square root of the direct sum of the honeycomb-lattice model with a sublattice-imbalanced on-site potential and the breathing kagome-lattice model, the latter of which is a representative model for the HOTI. 
We indeed find that the HOTI in the breathing kagome-lattice model is succeeded to this model,
as manifested by the existence of the corner states at finite energies and the nontrivial bulk polarization. 
The viewpoint of square-root operation is essential for the model construction.
Constructed as such, the parent and child Hamiltonians are indeed tied to each other in various aspects, 
such as bulk band structures, a topological invariant, and proper shape of boundaries to obtain the corner states.

The present construction of the square-root HOTIs will be applicable to various lattice models, 
including both two-dimensional and three-dimensional ones. 
For instance, for the decorated diamond lattice in three dimensions, its squared Hamiltonian is composed of the diamond and pyrochlore lattices, the latter of which hosts three-dimensional HOTIs~\cite{Ezawa2018}.
Exploring various other examples is an intriguing future work.

Seeking experimental realization of the square-root HOTI and the corner states in decorated honeycomb systems
is another interesting future problem. 
As we have emphasized, implementation of the decorated honeycomb-lattice structure in photonic crystals has been reported~\cite{Yan2019,Tang2020},
which will serve as a promising platform.
Indeed, our model of the Eq.~(\ref{eq:Ham_dh}) only includes two independent NN hoppings parameters, indicating that the model is feasible without any fine-tuning of parameters.
To realize the HOTI, the imbalance of two hoppings, $t_1 \neq t_2$, is essential, and such 
a situation can be realized by placing the sites on the edges of hexagons 
closer (or farther) to the sublattice (I) than the sublattice (II).  
Under such a setup, we expect that characteristic localized dynamics associated with the corner states can be observed. 
In addition, the nontrivial bulk polarization in the square-root HOTI may be experimentally accessible by observing the beam propagation if one can prepare a well-tailored beam profile as an initial state~\cite{Longhi2019}.

\acknowledgments
This work is supported by the JSPS KAKENHI, 
Grant No. JP17H06138 and No. JP20K14371 (T. M.), Japan.


\begin{thebibliography}{1}
\bibitem{remark_bk}
The breathing kagome lattice is a modified kagome lattice where upward triangles and downward triangles have different size.
Here by the breathing kagome-lattice model we mean the hopping on upward triangles is different from those on downward ones.
\bibitem{remark2}
Due to the chiral symmetry of the model, the sign of $t_{1/2}$ can be flipped by the gauge transformations.
\bibitem{remark3}
In fact, the probability distribution for the negative-energy and positive-energy modes are exactly the same, 
since these two are related by the chiral operation that does not change the amplitudes.
\bibitem{remark_4}
Note that this sum-rule-like relation
does not hold for the Berry connection, 
$A_{n,\mu}(\bm{k}):= -i \bm{u}_n^{\dagger}(\bm{k}) \cdot \frac{\bm{u}_n(\bm{k})}{\partial k_\mu}$, 
as one can check by using Eqs.~(\ref{eq:u5}) and (\ref{eq:u1}).
\end{thebibliography}
\end{document}